\documentclass[12pt,preprint]{aastex}
\usepackage[usenames,dvips]{color}

\newcommand{\lya}{Ly$\alpha$}
\newcommand{\feii}{Fe\,{\footnotesize II}} 
\newcommand{\cii}{C\,{\footnotesize II}} 
\newcommand{\ciii}{C\,{\footnotesize III}]} 
\newcommand{\civ}{C\,{\footnotesize IV}} 
\newcommand{\hi}{H\,{\footnotesize I}} 
\newcommand{\siii}{Si\,{\footnotesize II}} 
\newcommand{\siiv}{Si\,{\footnotesize IV}} 
\newcommand{\alii}{Al\,{\footnotesize II}} 
\newcommand{\znii}{Zn\,{\footnotesize II}} 
\newcommand{\crii}{Cr\,{\footnotesize II}} 
\newcommand{\mgi}{Mg\,{\footnotesize I}} 
\newcommand{\mgii}{Mg\,{\footnotesize II}} 
\newcommand{\invmic}{$\mu m^{-1}$}

\def\lax{{$\mathrel{\hbox{\rlap{\hbox{\lower4pt\hbox{$\sim$}}}\hbox{$<$}}}$}}
\def\gax{{$\mathrel{\hbox{\rlap{\hbox{\lower4pt\hbox{$\sim$}}}\hbox{$>$}}}$}}

\slugcomment{Submitted to ApJ \today}
\shorttitle{A metal-strong and dust-rich DLA} \shortauthors{Wang et
al.}

\begin{document}

\title{A Metal-Strong and Dust-Rich Damped \lya~Absorption System toward the Quasar SDSS J115705.52+615521.7}

\author{Jian-Guo Wang\altaffilmark{1,2,3,4},
Hong-Yan Zhou\altaffilmark{2,5},Jian Ge\altaffilmark{6},
Peng Jiang\altaffilmark{2,*}, Hong-Lin Lu\altaffilmark{2}, J. Xavier
Prochaska\altaffilmark{7}, Fred Hamann\altaffilmark{6},
Hui-Yuan Wang\altaffilmark{2},Ting-Gui Wang\altaffilmark{2} and
WeiMin Yuan\altaffilmark{8} }

\altaffiltext{1}{National Astronomical Observatories/Yunnan
Observatory, Chinese Academy of Sciences, P.O. Box 110, Kunming,
Yunnan 650011, China;~wangjg@ynao.ac.cn} \altaffiltext{2}{Key
Laboratory for Research in Galaxies and Cosmology, Department for
Astronomy, the University of Sciences and Technology of China,
Chinese Academy of Sciences, Hefei, Anhui 230026, China;}
\altaffiltext{3}{Key Laboratory for the Structure and Evolution of
Celestial Objects, Chinese Academy of Sciences}
\altaffiltext{4}{Graduate School of the Chinese Academy of Sciences,
19A Yuquan Road, P.O. Box 3908, Beijing 100039, China}
\altaffiltext{5}{Polar Research Institute of China, Jinqiao Rd. 451,
Shanghai, 200136, China} \altaffiltext{6}{Astronomy Department,
University of Florida, 211 Bryant Space Science Center, P. O. Box
112055, Gainesville, FL 32611, USA} \altaffiltext{7}{Department of
Astronomy and Astrophysics, UCO/Lick Observatory, University of
California, 1156 High Street, Santa Cruz, CA 95064, USA}
\altaffiltext{8}{National Astronomical Observatories, Chinese Academy of Sciences, Beijing 100012, China}
\altaffiltext{*}{LAMOST Fellow}

\begin{abstract}
We report the discovery of an unusual, extremely dust-rich and metal-strong
damped \lya\ absorption system (DLA) at a redshift $z_{a}=2.4596$ toward
the quasar SDSS J115705.52+615521.7 (hereafter J1157+6155) with an
emission-line redshift $z_{e}=2.5125$.
The quasar spectrum, taken in the Sloan Digital Sky Survey (SDSS),
shows a very red color and a number of metal absorption lines,
including \cii, \alii, \siii, \feii\, and \znii,
which are confirmed and further characterized by follow-up spectroscopy made with
the Multiple Mirror Telescope (MMT). Its neutral
hydrogen column density $N_{\hi} = 10^{21.8\pm0.2}$~cm$^{-2}$ is
among the highest values measured in quasar DLAs. The measured metal
column density is $N_{ZnII}\approx 10^{13.8}$~cm$^{-2}$, which is
about 1.5 times larger than the largest value in any previously
observed quasar DLAs. We derive the extinction curve
of the dusty DLA using a new technique, which is an analog of the
``pair method'' widely used to measure extinction curves
in the Milky Way (MW). The best-fit curve is a MW-like law
with a significant broad feature centered around 2175~{\AA} in the
rest frame of the absorber. The measured extinction $A_V \approx 0.92$ mag is
unprecedentedly high in quasar DLAs. After applying an extinction
correction, the $i$ band absolute magnitude of the quasar is as high
as $M_{i} \approx -29.4$ mag, placing it one of the most luminous
quasars ever known. The large gas-phase relative abundance of
[Zn/Fe] $\approx1.0$ indicates that metals are heavily depleted onto
dust grains in the absorber. The dust depletion level is between
that of the warm and cool clouds in the MW. This discovery is suggestive
of the existence of a rare yet important population of dust-rich DLAs with
both high metallicities and high column densities, which may have
significant impact on the measurement of the cosmic evolution of
neutral gas mass density and metallicity.

\end{abstract}


\section{Introduction}
High luminosity quasars are a very useful probe into the early
Universe. Gas intervening between us and a distant quasar will leave
imprints of absorption lines in the quasar spectrum.
DLAs\footnote{In this paper, this abbreviation
is referred to as quasar DLAs unless further specified.} are a
class of special quasar absorbers with high hydrogen column density,
$N_{\hi} \ge 10^{20.3}$~cm$^{-2}$, by definition (see Wolfe et al.
2005 for a comprehensive review). Because of the high column
density, gas in DLAs should be mainly neutral under the illumination
of background radiation from quasars and galaxies at high redshift
(e.g. Prochaska \& Wolfe 1996). DLAs are usually considered as the
main neutral gas reservoir for star formation in the high redshift
Universe (e.g. Nagamine, Springel \& Hernquist 2004; Prochaska,
Herbert-Fort \& Wolfe 2005).

For a quasar at a redshift $z$, its absorption distance (Bahcall \&
Peebles 1969) is defined as
\begin{equation}
X(z) = \int_0^z \frac{(1+z')^2}{\sqrt{(1+z')^2(1+z'\Omega_m) -
z'(z'+2)\Omega_{\Lambda}}}dz'.
\end{equation}
In principle, we may calculate the mean co-moving H~I density by
averaging $N_{\hi}$ in different sight lines,
\begin{equation}
\Omega_{\hi}(X)dX = \frac{\mu
m_HH_0}{c\rho_c}\int_{N_{min}}^{\infty} N_{\hi}f(N_{\hi},X)dN_{\hi},
\end{equation}
where $\mu = 1.3$ is the mean molecular weight of absorption gas,
$m_H$ the hydrogen atomic mass, $H_0$ the Hubble constant, $\rho_c$
the critical mass density, and $f(N_{\hi},X)$ the neutral hydrogen
column density distribution of quasar absorbers. The observed $f(N_{\hi},X)$
can be crudely described either by a $\Gamma$ function (Fall \& Pei 1993;
P{\'e}roux et al. 2003) or by a double power-law with a break at
$\log (N_{\hi}/cm^{-2}) \sim 21.5$ (Prochaska \& Wolfe 2009;
Noterdaeme et al. 2009).

Although great efforts have been made to measure the neutral hydrogen column
density, $N_{\hi}$, and frequency distribution,  $f(N_{\hi},X)$, of DLAs,
the slope at the high $N_{\hi}$ end is not well constrained
by observations. The slope varies from $\alpha \sim -6.0$ (Prochaska \& Wolfe
2009) to $\alpha \sim -3.5$ (Noterdaeme et al. 2009), unlike the stable measurement of slope at the low
$N_{\hi}$ end, $\alpha \sim -2$ (P{\'e}roux
et al. 2005; Prochaska \& Wolfe 2009; Noterdaeme et al. 2009).  Such a
disagreement is likely originated from the large statistical fluctuation caused by the rare detections of high $N_{\hi}$ DLAs.
Since the mean co-moving H~I density, $\Omega_{\hi}(X)$ (Bahcall \& Peebles 1969),
depends sensitively on the slope of
$f(N_{\hi},X)$ at high $N_{\hi}$\footnote{$f(N_{\hi},X)$ is extrapolated
to infinity when taking the integral in Equation (2).},
the uncertainty (of the slope) leaves
the measurement of $\Omega_{\hi}(X)$ inaccurate. It is quite desirable to measure
more high column density DLAs to determine the shape of $f(N_{\hi},X)$
at the high $N_{\hi}$ end correctly. However, so far, only two
super-DLAs with $\log (N_{\hi}/cm^{-2}) \sim 22$ have been
identified, SDSS J081634.40$+$144612.9 (hereafter J0816$+$1446;
Noterdaeme et al. 2009; Guimar{\~a}es et al. 2012)
and SDSS J113520.39$-$001052.5 (hereafter J1135$-$0010; Kulkarni et al. 2012;
Noterdaeme et al. 2012).

The paucity of high $N_{\hi}$ DLA detections can be attributed
either to their intrinsic rarity or to selection effect induced
by dust obscuration or both. Early evidence for dust in DLAs was found by
comparison of the continuum slope between quasars with and without
DLAs (e.g. Fall \& Pei 1989; Fall, Pei \& McMahon 1989; Pei, Fall \&
Bechtold 1991), which was later confirmed by the studies of the
differential depletions between refractory and volatile elements
(e.g. Pettini et al. 1994, 1997; Vladilo et al. 2002).
Currently, only two dust-rich DLAs ($A_V \approx 0.2$ in
SDSS J0918+1636, Fynbo et al. 2011; and $A_V \approx 0.4$ in AO
0235+164, Junkkarinen et al. 2004) have been reported.
On the contrary, dusty absorbers are detected frequently in other astrophysical
environments. In the MW, dusty clouds with $A_V \sim 1.0$ mag
are commonly seen (Diplas \& Savage 1994). High dust
extinction was measured in the local galaxies as well (e.g. Xiao
et al. 2012). Even at high redshift, many very dusty absorbers have
been detected by observations of GRB afterglow (e.g. Schady et al. 2011).
Fall \& Pei (1993) developed an analytical model to study the selection effect
due to the obscuration of dust in DLAs and suggested that a fairly large fraction of
dusty DLAs has been overlooked in the quasar surveys.
The full obscuration model requires a luminosity function of
quasars, a distribution of dust-to-gas ratio and a distribution of
$N_{\hi}$ in DLAs. The application of the obscuration model is
presented in \S5.

Star formation leads to consumption of neutral gas and
enrichment of heavy elements. The measurements of average
metallicity at different redshifts provide robust diagnosis for
cosmic star formation. The DLA technique could be a powerful tool to
probe the metals in the high redshift Universe. However, current
measurements suggest that DLAs are generally metal-poor. The mean metallicity
is only a few percents of the solar abundance (e.g., Kulkarni \& Fall 2002;
Prochaska et al. 2003; Kaplan et al. 2010). It is significantly
lower than the metallicity measured in the MW and the nearby galaxies
as well as the high redshift galaxies selected through their starlight
emissions, such as Lyman break galaxies (e.g., Giavalisco 2002 and references therein).
Since dust grains are made of heavy elements, the correlation between metallicity
and dust abundance is expected (e.g., Xiao et al. 2012). Therefore,
a direct explanation for the deficiency of heavy elements
in DLAs is that the metal-rich absorbers suffer relatively large
dust obscuration and thus are systematically omitted in the DLA
surveys (e.g., Boiss\'{e} et al. 1998).

In this paper, we report the discovery of an unusual DLA with high
column density $N_{\hi} = 10^{21.8\pm0.2}$~cm$^{-2}$ toward the
quasar J1157+6155. It is the
metal-strongest and dust-richest quasar DLA to the best of our
knowledge. We further characterized the property of its dust content
by evaluating the extinction curve in the absorption rest frame. The
resulted curve exhibits a significant 2175~{\AA} extinction bump,
which is very rarely seen in quasar absorption line systems (e.g.,
Wang et al. 2005; Noterdaeme et al. 2008; Zhou et al. 2010;
Jiang et al. 2011). The paper is organized
as follows. In \S2, we describe the data used in this paper,
including archived data and our MMT spectroscopic data. We evaluate
the extinction curve in \S3 and measure the absorption line spectrum
in \S4. The results are summarized and discussed in \S5. Throughout
this paper, we adopt the cometic abundances from Asplund et al.
(2009) as solar abundances, and assume a cosmology with $H_0= 70$~km~s$^{-1}$
Mpc$^{-1}$, $\Omega_{\rm M}=0.3$, and $\Omega_{\Lambda}=0.7$.

\section{Observations and Data Reduction}
J1157$+$6155 was imaged in the SDSS on May 18, 2001 in the $u$, $g$,
$r$, $i$, and $z$ bands, giving point-spread-function (PSF)
magnitudes of 24.58$\pm$1.23, 20.69$\pm$0.03, 19.69$\pm$0.03,
19.24$\pm$0.02, and 18.47$\pm$0.04 (AB-system), respectively.
J1157$+$6155 was selected as a high-redshift quasar candidate based
on its location in the `$griz$ color' cube (Richards et al. 2004),
and was confirmed as a quasar at an emission line redshift $z_e =
2.5125 \pm 0.0013$ by an spectroscopic observation performed on Feb.
15, 2002. The SDSS spectrum, extracted from the SDSS data release 7
(DR7; Abazajian et al. 2009), is very red with a number of
absorption lines imposed. We identified a broad absorption feature in the
blue end of the spectrum as a DLA at the same redshift as those of
the metal absorption lines ($z_a = 2.4596$). Of particular interest are very strong
absorption features at the $\lambda $2026 \znii\ +\mgi\ blend and
the $\lambda $2062 \znii\ +\crii\ blend with rest-frame equivalent
widths (EW) of $\sim 1$ \AA, which have never been seen in any
previous quasar DLAs, as far as we know.

To investigate J1157$+$6155 in more detail, we performed follow-up
spectroscopic observations using the Blue Channel Spectrograph mounted
on the 6.5~m MMT. Two exposures with length of 1200 s were taken by
using the 500~g~mm$^{-1}$ grating on February 11, 2008. Another exposure
with length of 900 s was obtained by using the 800~g~mm$^{-1}$ grating
on March 30, 2008. A slit width of $1^{''}.0$ was chosen to match the
seeing in both of the nights.
The 500~g~mm$^{-1}$ grating was blazed at
6000 \AA\ (1734 \AA\ in the DLA rest frame) and the 800~g~mm$^{-1}$
grating at 4000 \AA\ (1156 \AA~ in the DLA rest frame). They
provide a wavelength coverage of $\lambda \sim 4400 - 7500$ \AA\ (1272
- 2168 \AA\ in the DLA rest frame) and $\lambda \sim 3200 - 5200$
\AA\ (925 - 1503 \AA\ in the DLA rest frame), respectively. The
corresponding spectral resolutions are 3.8 and 2.5 \AA~in full width
at half maximum (FWHM) measured from the arc lamp lines. The
CCD reductions, including bias subtraction, flat$-$field correction,
and cosmic ray removal, were accomplished with the standard procedures
using IRAF\footnote{IRAF is distributed by the National Optical
Astronomy Observatory, which is operated by the Association of
Universities for Research in Astronomy, Inc., under cooperative
agreement with the National Science Foundation.}. Wavelength
calibration was carried out using He/Ne/Ar lamp. A KPNO standard
star was observed for flux calibration.

J1157$+$6155 was also detected on March 1, 1993 during the Two Micron
All Sky Survey (2MASS; Skrutskie et al. 2006) with magnitudes (Vega-system) of
$16.894 \pm 0.183$, $16.314 \pm 0.202$ and $15.093 \pm 0.155$ in the
$J$, $H$ and $K_s$ band, respectively. All of the
photometric and spectrophotometric data were first corrected for
Galactic reddening using the extinction map of Schlegel et al.
(1998) and the reddening curve of Fitzpatrick (1999) before performing further
analysis. The three MMT spectra, after scaled by a small factor to
compensate for the aperture effect and the uncertainty of absolute
flux calibration, are consistent with the SDSS spectrum in the
overlapped wavelength range. The SDSS and MMT spectra are combined
to form one spectrum covering a wavelength range of $3200 - 9200$\AA\
($925 - 2660$ \AA\ in the DLA rest frame).
The combined spectrum is used to derive an extinction
curve in \S3.

\section{Reddening and Extinction Curve}

In this section, we explore the distribution of the spectral indices
of quasars and derive the extinction curve of the DLA with two methods.

\subsection{Distribution of Quasar Spectral Indices}

The rest-frame UV to optical spectral energy distribution (SED) of
the quasar J1157$+$6155 is extremely red, suggesting that
it is heavily reddened by the foreground DLA.
To gauge the dust extinction of
J1157$+$6155, we first examined the distribution of spectral indices
$\alpha$ of SDSS quasars in the redshift interval $2.4 < z <
2.6$, which is defined as $S_{\lambda}\propto \lambda^{\alpha}$ and
calculated in two continuum windows (1690--1720\AA\ and
2225--2250\AA) in the quasar rest frame. The distribution function
is roughly a Gaussian with a red ``tail'' (see Fig. 1). The red
``tail'' is similar to the red ``tail'' of the color distribution of
quasars which was attributed to dust reddening (Hopkins et al.
2004).
The spectrum of J1157$+$6155 is very red, with an index $\alpha
\approx0.24$, that deviates significantly
from the vast majority of the distribution.
This motivates us to model its extinction curve
in detail.

\subsection{Overview of the Modeling Methods}

Light from a quasar behind an intervening DLA is subject to
wavelength-dependent extinction by dust therein. The extinction
curve can be characterized by comparing the observed spectra with
the intrinsic one, which are usually unknown. A composite spectra of
quasars, scaled by a factor, is often used as a substitute to the
intrinsic spectra (e.g. Wang et al. 2004; Zhou et al. 2010).
However, the feasibility of using
composite quasar spectra is questionable because of the rich
diversity of quasar spectra. In order to alleviate this problem, we
design a new method to derive extinction curves and to gauge the
uncertainties induced by mismatches among the intrinsic quasars
spectra. We first build up a spectra library of quasar  as templates
to model the intrinsic spectrum of a target. The observed target
spectrum is then fitted with models as the template spectra drawn
from the library that are reddened by dust with a parameterized
extinction curve. These fittings with different template spectra
result in distributions of the best-fit parameters, which are in
turn used to derive the best-estimates of the parameters and their
systematic uncertainties due to mismatches among the templates. We
refer to this method as ``quasar spectra pair method''.

\subsection{Composite Quasar Spectrum Method}

A parameterized method for determining extinction curves was
described by Zhou et al. (2010) and Jiang et al. (2010a,b; 2011).
The spectrum of a target is modeled by a
composite quasar spectrum, which is reddened
with a parameterized extinction curve. The
best-fit parameters are determined by minimizing $\chi^2$ using
MPFIT (Markwardt 2009). We use the
parameterized UV/optical extinction curve of Fitzpatrick \& Massa (1990, 2007),
\begin{equation}
k(\lambda) = c_1 + c_2 x + c_3 D(x,x_0,\gamma)
\end{equation}
where $x \equiv \lambda^{-1}$, in units of inverse microns
(\invmic). There are five
free parameters in the formula which
correspond to two features in the curve: (1) a linear component
underlying the entire UV wavelength range, with two parameters $c_1$ and
$c_2$; (2) a Lorentzian-like 2175~{\AA} bump (e.g., Tielens 2005),
described by three parameters: strength $c_3$, center wavelength $x_0$, and
width $\gamma$, then expressed as
\begin{equation}
D(x,x_0,\gamma) = \frac{x^2}{(x^2-x_0^2)^2 +x^2\gamma^2} \;\; .
\end{equation}

Following Zhou et al. (2010), the composite quasar spectrum is
obtained by combining the SDSS composite spectrum ($\lambda \leq 3000 \AA$;
Vanden Berk et al. 2001) and the near-infrared template ($\lambda >
3000 \AA$; Glikman et al. 2006).
In the fitting procedure, only the wavelength range longer than
1250 \AA\ is used, and small
weights are assigned to regions around strong broad emission lines,
including \mgii$\lambda2800$, \ciii$1909$, \civ$1549$ and
\siiv$1400$, while absorption lines are masked.
The best-fit results are $c_1 = -2.75 $, $c_2 = 0.48$, $c_3 = 0.43$, $x_0 =
4.82$, and $\gamma = 1.13$. The best-fit model (red line)
is overplotted with the observed spectrum in the upper panel of Fig. 2.
In the near-infrared band, the resultant model is consistent with 2MASS photometry.
We also attempted to fit the observed spectrum using an SMC-like extinction
curve (Pei 1992; Gordon et al. 2003). The resultant spectrum disagrees with
the observed one around 2000 -- 2200 \AA\ in the DLA rest frame, indicating
the requirement of an absorption bump.
In the bottom panel of Fig. 2, we compare the SMC-like extinction curve (dotted black line) with
the best-fit extinction curve (solid black line).
The extinction curve derived here is subject to an arbitrary shift (on $c_1$)
of the composite quasar spectrum (Zhou et al. 2010; Jiang et al. 2011).\footnote{
$c_1$ depends on the flux level of the composite quasar spectrum, which is arbitrary
selected.} In fact, a physical extinction curve needs to be zero when it goes toward
$\lambda \rightarrow \infty$. Thus, we can physically determine the parameter $c_1$ by
forcing the extinction to be zero when
extrapolating the best-fit extinction curve to $\lambda \rightarrow \infty$.
It yields $c_1 = 0$ and $A_V = 0.88$ mag.

\subsection{Quasar Spectra Pair Method}

\subsubsection{Building up the Template Library}

Composite quasar spectra were often used as a substitute to the
intrinsic spectra of target quasars in previous work.
However, as aforementioned,
the feasibility of such substitution is suspect because
of the rich diversity of the quasar continuum and pseudo-continuum
composed of a large number of \feii\ emission lines (Pitman et al.
2000). To explore
the systematic uncertainty due to the diversity of quasar spectra,
we build up a library of quasar spectra, and use them to model the
intrinsic spectrum of target quasars.
In this process the parameters of the extinction curves are extracted.
The distributions of derived parameters are then used to
gauge the systematic uncertainties introduced by the spectral diversity.

The library of quasar spectra in the redshift range of
$0.6 < z < 2.4$, which are selected from the SDSS DR7 quasar catalog (Schneider
et al. 2010), is built up in the following steps. Note that
we do not build templates with $z_e \ge 2.4$ because the number of quasars
with high quality SDSS spectra decreases quickly beyond this redshift.

\begin{enumerate}
\item{ Discarding possible Broad absorption line (BAL) quasars by
   making use of the catalogs of Gibson et al. (2009),  Scaringi et al.
   (2009) and Shen et al. (2011).}
\item{ Calculating the power-law index of quasar continuum.
   The power-law index is calculated using
   the continuum windows: $1690-1700$ \AA~and $2225-2250$ \AA~for quasars
   with redshifts $z \ge 2$, $2225-2250$ \AA~and $3030-3090$ \AA~for quasars
   with redshifts $0.8<z<2$, $3030-3090$ \AA~and $4020-4050$ \AA~for quasars
   with redshifts $z \le 0.8$.}
\item{ Selecting objects according to the distribution of the power-law indices.
   In each redshift interval of $\delta z = 0.2$, the distribution of the power-law
   indices is found to be roughly a Gaussian with a ``red tail'', as shown in Fig. 1.
   Quasars in the ``red tail'' (to the right of the $2\sigma$ dashed line)
   may be either intrinsically red or reddened by dust.
   We assume that the un-reddened intrinsic
   distribution of the indices is Gaussian.
   We randomly select quasars from the ``red tail'' of the distributions
   while keep the others (on the left of
   the $2\sigma$ dashed line), so as to make the resulting
   distributions of the indices following a Gaussian distribution
   (similar to the red line in Fig. 1).
   The first $\sim 500$ quasars with the highest S/N ratios are selected in each
   redshift interval.}
\item{ Removing narrow absorption lines from the selected spectra.
Each of the spectra selected (denoted as $f_0$)
is smoothed using the robust
local regression smoothing (Cleveland 1979; denoted as $f_1$), and then
the initial spectrum
is normalized by its smoothed spectrum, $f_2=f_0/f_1$.
For $f_2$, pixels
with values smaller than 0.85 may be part of absorption
lines and their values are set to 1.
Narrow absorption lines superimposing
over broad emission lines, such as \mgii~$\lambda$2800,
\ciii~$\lambda$1909, \civ~$\lambda$1549 and \siiv~$\lambda$1396,
cannot be effectively recovered with this approach.
The profile of each
of the broad emission lines in the normalized spectrum $f_2$ is then
fitted by a Gaussian. The fitting result is denoted as $f_3$.
$f_2$ is then re-normalized by $f_3$, $f_4=f_2/f_3$.
Similarly, pixels of $f_4$  with values smaller than 0.85
are set to 1.
The ``true''
spectrum of a quasar is the product of $f_4\times f_3\times f_1$.}
\end{enumerate}
Finally, all the template spectra are visually inspected to ensure
that there are no apparent absorption lines.

\subsubsection{Refined Modeling}
After the library of the intrinsic spectra of quasars is built up,
the target spectrum is fitted with the $\sim$500 template spectra in
a redshift interval $[z_e, z_e+0.2]$. For J1157$+$6155, the templates
are obtained in the redshift interval $[2.2, 2.4]$, since the high
redshift template with $z_e > 2.4$ is not available in the library.
Only the common wavelength range of the
target spectrum and the template spectra are used in the fit, which
is $ 1250 - 2620$ \AA\ for J1157$+$6155 in the quasar rest frame.
Next, we will extract the intrinsic strength of \feii\ emission of
quasar J1157$+$6155 and constrain the extinction curve fitting with only
the template quasar spectra having a similar \feii\ emission strength.
The refined modeling procedure can help to reduce the contamination
from \feii\ emission when measuring the bump profile.

We first take the median of each of the parameters of the extinction
curve from fitting the $\sim 500$ templates as the best-fit values,
and de-redden the observed spectrum with the median extinction curve.
Then we model the \feii\ emission of the de-reddened spectra using the method
described in Wang et al. (2009) and estimate its EW. The \feii\ template
used here is a combination of the \feii\ template of Vestergaard \& Wikes
(2001; $\lambda < 2200$~\AA) and that of Tsuzuki et al. (2006; $\lambda \geq 2200$~\AA).
Next we measure the \feii\ EWs for every quasar in the library in the same way.
Only the quasars whose \feii\ EWs are consistent with
that of the target spectrum within $1\sigma$ uncertainty are
selected as the templates for refined modeling. This selection results in $\sim 100$
\feii\ matched template spectra. We repeat the extinction curve fitting
procedure with the \feii\ matched template spectra. Finally, the best-fit
parameters and their uncertainties are adopted from the distributions of the
fitted parameters, which can be modeled with Gaussians (Fig. 3).
They are $c_2 = 0.50\pm 0.04$, $x_0 = 4.79\pm 0.04$,
$\gamma = 0.98\pm0.20$ and $c_3 = 0.37\pm 0.18$, where the values
are the expectations and the errors are the standard deviations.
Since $c_1$ is an arbitrary parameter depending on the brightness of
template spectra as stated above, it is not presented in Fig. 3.

The significance of the 2175~{\AA} absorption bump can be gauged
by the distribution of bump strength (Jiang et al. 2011), which is
measured by the area of the bump $A_{bump} = \pi c_3/2\gamma$.
The red Gaussian (in panel $e$ of Fig. 3) is the distribution of bump strength
derived with the \feii\ matched template spectra. The expectation is
$A_{bump} = 0.61$ and the standard deviation is $\sigma = 0.19$. Thus,
the null hypothesis (i.e. no 2175~{\AA} bump or $A_{bump} = 0$) can
be rejected at a statistical confidence level greater than $3\sigma$.
Alternatively, the 2175~{\AA} absorption bump in J1157$+$6155 is
detected at a confidence level of $>3 \sigma$.
In the same panel, we also present the bump strength distribution
derived with the full template library (the blue Gaussian), which
is nearly twice broader than that derived with the \feii\ matched template spectra.
In this case, the 2175~{\AA} absorption bump is only detected at a
confidence level of $2 \sigma$.
The comparison clearly shows the contamination of the variations of
\feii\ broad emissions on the measurements of bump strength (Pitman et al. 2000).
We conclude that the refined modeling procedure is necessary for
detecting relatively weak 2175~{\AA} absorption bumps on quasar spectra.
Given the \feii\ emissions on the composite quasar spectrum might not
match with that on the observed spectrum, the quasar spectra pair method
should be more accurate than the composite quasar spectrum method on measuring
the 2175~{\AA} absorption bumps in general.
The comparison of extinction curve derived with quasar spectra pair method
and that of the composite quasar spectrum method is presented in Fig. 4.

The parameter $c_1$ is recalculated by forcing the extinction to zero
for $\lambda \rightarrow \infty$. This gives  $A_V= 0.92 \pm
0.07$ mag and a total-to-selective extinction ratio of $R_V \equiv
A_V/E(B-V)\approx 3.97$. The slope of the underlying extinction is
similar to that found in J1007+2853 ($R_V\approx 3.87$), a very
dusty quasar with super-strong 2175~{\AA} absorption (Zhou et al.
2010). Both slopes are larger than the averaged value of $R_V =
3.1$ in the MW (e.g., Draine 2003). The background quasar is
intrinsically very bright with $i\approx 16.5$ mag after
de-reddening using the derived extinction curve; this corresponds to
an intrinsic luminosity of $M_i\approx -29.4$ mag, assuming an
optical spectral slope of $\alpha_{\nu}=0.5$ ($S_{\nu} \propto
\nu^{-\alpha_{\nu}}$; Schneider et al. 2010). We note that the 2MASS $K_s$ band
absolute magnitude is as high as $M_{K_s}\approx -31.5$ even without
extinction correction\footnote{We assume an optical to near-infrared slope of
$\alpha_{\nu}=1$ ($S_{\nu} \propto \nu^{-\alpha_{\nu}}$; Glikman et
al. 2006).}. In addition, the redshift of 2175~{\AA} bump in J1157$+$6155
$z=2.46$ is slightly higher than the previously highest redshift 2175~{\AA}
absorber toward the Gamma Ray Burst (hereafter GRB) 070802 ($z=2.45$;
El{\'{\i}}asd{\'o}ttir et al. 2009).

\section{Column Densities, Gas-Phase Abundances and Dust Depletion}

The \lya\ absorption line of J1157$+$6155 is covered by both SDSS and MMT spectra.
We perform line profile fitting on the MMT spectrum, since it has
higher signal-to-noise ratio. The fitting is realized by using the
program x$\_$fitdla from the XIDL
package\footnote{http://www.ucolick.org/$\sim$xavier/IDL}. The best-fit
Voigt profile, as shown in Fig.5, yields a column density
$\log N_{\hi} = 21.8 \pm 0.2$~cm$^{-2}$. There are additional absorptions
blueward of the DLA, which may be attributed to unidentified absorption lines
corresponding to absorbers at different redshifts.

In order to measure the column density of metal absorption lines, we normalize
the observed spectrum using an absorption-free quasar continuum modeled by a series
of polynomials. The normalized spectra are plotted in velocity space in
Fig. 6. The Fe$\lambda\lambda$ 2344, 2374, 2382, 2586, 2600 absorption lines
are taken from SDSS spectrum, because they are not covered by our MMT spectra.
The curve of growth (COG) method is employed to measure the absorption
line column densities of heavy elements (Jenkins 1986).
The equivalent widths (EWs) of all transitions are calculated by integrating
over an velocity interval [-800,400]~km~s$^{-1}$, which is
so-chosen to cover most of the
absorption of the detected transitions and void the contamination
from nearby absorptions. The \feii\ $\lambda\lambda$ 1608,
2344, 2374, 2382, 2586 and 2600 absorption lines, arising from
the ground level of $Fe^+$, are used to derive the COG.
The best-fit column density $N_{\feii}=15.62\pm0.19$~cm$^{-2}$
and Doppler parameter $b=182\pm26$~km~s$^{-1}$, are obtained by
$\chi^2$ minimization. We further explore the $\chi^2$ space by stepping the
$N_{\feii}$ and $b$ to construct the contours of the 68\% and 90\%
confidence levels. The error-bars reported above are corresponding to
the 68\% confidence interval. Using the best-fit COG (Fig. 7),
we derive the column densities of other ions. For species having multiple
transitions available, the column densities are obtained by fitting all the absorptions
simultaneously.

The derived Doppler parameter $b$ from COG is relatively large. It may indicate that
the absorption lines have narrow sub-structures, which cannot be resolved on
the MMT and SDSS spectra. Thus, the column densities based on the single-component
COG might be underestimated (Prochaska 2006).
We also measure the column densities of all the absorption lines using the apparent
optical depth method (AODM; Savage \& Sembach 1991). The resultant column
densities are systematically smaller than that derived by COG method. The difference
is expected since the absorption lines are heavily saturated generally. Thus, we adopt
the column densities on the basis of COG in the rest of this paper. The measurements
are summarized in Table 1.

The measured $N_{\znii}$ of J1157$+$6155 is $\sim$1.5 times higher than that of
the previous record holder, SDSS J1137$+$3907, whose
$N_{\znii} = (2.692 \pm 0.311) \times 10^{13}$~cm$^{-2}$ (Meiring et al. 2006).
Since the column density of neutral hydrogen is very large, the
ionization correction of metal elements is ignorable. Then, the
gas-phase abundances relative to solar values are derived, yielding
[Zn/H]\footnote{Chemical abundance relative to the solar
value defined as $[X/Y]\equiv \log(X/Y)-\log(X/Y)_{\odot}$.}
= $-0.60\pm0.38$, [Si/H] = $-0.83\pm0.27$, [Fe/H] = $-1.62\pm0.28$,
[Cr/H] $<-1.36$, [Ni/H] $< -1.32$. J1157$+$6155 is the metal-strongest DLA to the
best of our knowledge. In fact, these abundances might be even
larger, since the column densities of metal absorption may have been underestimated
as stated above.
Boiss\'{e} et al. (1998) noted that all DLAs in their sample lie
below an threshold ($N_{\znii} = 1.4 \times 10^{13}$ cm$^{-2}$) in
the [M/H]-$N_{\hi}$ plane, and explained that
the threshold was set by the dust obscuration bias. A similar trend has been
seen in a larger sample of DLAs compiled by Prochaska et al. (2007)
as well (see Fig. 8). The lack of metal-strong DLA in the previous
surveys may be caused by the disadvantage of conventional quasar
selecting methods on the basis of the UV-blue excess.
The SDSS uses a quasar selection criterion (Richards et al. 2002) that
is more compatible with dust reddened quasars than the conventional
methods. Herbert-fort et al. (2006) obtained a sample of metal-strong
DLAs having similar metallicity with the MW clouds (Roth \& Blades 1995)
in SDSS.\footnote{J1157$+$6155 in this work can be classified as a metal-strong
DLA according to the definition in Herbert-fort et al. (2006).}
Kaplan et al. (2010) studied the quasar color of this DLA sample
and found significant dust reddening. Among the metal-strong DLAs (filled circles in Fig. 8),
J1157$+$6155 has the highest column density of gas. The population of high
column density and metal-strong DLAs, bearing bulks of heavy elements,
has great weight on the measurement of the cosmic metallicity.

In Fig. 9, we show comparison between the dust depletion pattern of J1157$+$6155 and that
of the MW cool/warm disk clouds on the line of sight towards $\zeta$ Oph
(Savage \& Sembach 1996). The dust depletion pattern of J1157$+$6155 is
quite similar with the MW clouds. The MW-like depletion pattern has been
seen in two 2175~{\AA} quasar absorbers previously (Jiang et al. 2010b).
The heavy dust depletion confirms the existence of dust grains and
indicates that J1157$+$6155 may be a $H_2$-bearing DLA according to
the correlation between the presence of $H_2$ and high depletion level
(Ge \& Bechtold 1997; Ge et al. 2001; Noterdaeme et al. 2008).

Vladio et al. (2008) derived the average dust-to-gas ratio
of DLAs ($A_V/N_{\hi} \approx 3/10^{22}$ mag cm$^2$; the dotted curve
in Fig. 10), by comparing the color of 248 quasars with DLAs
and that of a large non-DLA quasar sample. The
dust-to-gas ratio of the three dust-rich DLAs (J0918$+$1636 and
AO 0235$+$164 as well as J1157$+$6155 in this work) is larger than
the average value by a factor of 5. While the two
super-DLAs (J0816$+$1446 and J1135$-$0010) have high column
densities of gas but very little dust content. Their dust-to-gas
ratio is about 1/4 of the average value of DLAs. We conclude that
the dispersion of dust-to-gas ratio in DLAs is quite large and can vary
over an order of magnitude.

\section{Discussion}

The uniqueness of J1157$+$6155 endows it an ideal laboratory to
explore DLA properties. Its neutral hydrogen column density of
$N_{\hi} = 10^{21.8\pm0.2}$~cm$^{-2}$ is the top several among known
quasar DLAs. The measured $N_{\znii}$ is the highest among quasar
DLAs to the best of our knowledge. Combining the SDSS spectroscopic
data and the 2MASS photometric data of J1157$+$6155, we inferred an
extinction as large as $A_V\approx0.92$~mag, which is much larger
than that ever reported in quasar DLAs. The presence of dust grains
is further confirmed by the high dust depletion levels measured on
our following up MMT spectrum. The discovery of DLA J1157$+$6155 may
reveal the existence of an important population of metal-strong and
dust-rich super-DLAs.

This new population of DLAs might be overlooked due to dust obscuration in
the previous DLA surveys, which are generally based on optically selected
quasar samples. Even if occasionally detected, the metal-strongest DLAs
tend to be found at the faint end of quasar brightness distributions
in magnitude-limited surveys.
Herbert-fort et al. (2006) compiled a sample of metal-strong DLAs and gauged
the significance of the correlation between $N_{Zn^+}$ and r-magnitude.
The correlation is not significant as expected by the authors.
We note that the censored data were excluded in their statistical study.
By taking the censored data
into account and adding two new metal-strong DLAs, J1157$+$6155 in this work
and the one at $z=2.58$ towards the quasar SDSS
J0918$+$1636 reported by Fynbo et al. (2011), we find a
marginal correlation between r-magnitude and $N_{Zn^+}$ with a
Spearman coefficient of $r_S=0.39$ and a chance probability of
$P_{null}< 3.9\%$ using the ASURV package (Isobe et al. 1986). It
suggests that DLAs, at least metal-strong DLAs, are likely to be affected by
dust obscuration. Attenuated by very dusty absorbers, even quasars
as intrinsically luminous as J1157$+$6155---one of the few ten
highest luminosity quasars in about one hundred thousand SDSS
quasars---could easily escape from quasar DLA surveys. Herbert-fort et al.
(2006) adopted a brightness requirement of $r<19.5$ mag when they
compiled the metal-strong DLA sample. With $r=19.69$ mag, J1157$+$6155 is
just escaped from this brightness threshold.

Therefore, the dust-rich DLAs should mainly be detected toward the faint
quasars. Deep quasar surveys are desired to disclose
the abundance of dust-rich DLAs with both high metallicities and high column
densities. Since only intrinsically luminous quasars can be observed after
intersecting a very dusty absorber, an amplification factor on their
number density is necessary to correct the selection effect induced
by dust obscuration in magnitude limited quasar surveys.
We estimate the ratio of the real
number density and the observed number density of dusty DLAs having a similar
optical depth and redshift as J1157$+$6155 on the basis of the dust obscuration
model built by Fall \& Pei (1993). By defining $\rho_{t}(\tau, z)$ as the
mean numbers of absorbers along random lines of sight with optical
depths $\tau$ at a fixed redshift $z$ and $\rho_{o}(\tau, z)$ as the
mean numbers along observed lines of sight, the correction factor
can be expressed as:
\begin{equation}
\eta = \rho_{t}(\tau, z) / \rho_{o}(\tau, z) = e^{\tau \beta},
\end{equation}
where $\tau$ is the optical depth in $V$ band at $z=0$,
and $\beta$ is the negative exponent at the bright end of the quasar
luminosity function. For simplicity, we assume that the extinction
law is $\tau_{rest} \propto 1/\lambda$ in the DLA rest frame. Then the
optical depths $\tau$ in the equation (5) yields
$\tau = \tau_{rest}(1+z) \sim A_{V}^{rest}(1+z)$. Applying the
measured $A_V^{rest}$ and $z$ in the case of J1157$+$6155 and $\beta
\sim 2.1$ from the SDSS quasar luminosity function (Richards et al.
2002), we obtained a huge correction factor, $\eta \sim 1,000$.
One serendipitous detection of the dust-rich DLAs in the $\sim 12,000$
high redshift ($z > 2.5$) quasars of SDSS DR7 (Schneider et al. 2010)
and the huge amplification factor of observed number density jointly indicate
a significant population of dust obscured DLAs.\footnote{It is beyond the scope
of this paper to measure the number density.}

The cosmological mean metallicity
$\langle Z \rangle$ (i.e., the $\log$ of the ratio of the co-moving densities of
metals and gas relative to the solar abundance) measured in DLAs can be calculated as:
\begin{equation}
\langle Z\rangle = \log \left( \frac{\int_{0}^{\infty} N_Zf(N_Z)dN_Z}{\int_{N_{min}}^{\infty} N_{\hi}f(N_{\hi})dN_{\hi}}\right) - \log (N_Z/N_{\hi})_{\sun},
\end{equation}
where $N_Z$ and $N_{\hi}$ are the column densities of metals and neutral hydrogen gas.
Affected by the dust obscuration bias, nearly all the known DLAs lie below the
threshold ($N_{\znii} = 1.4 \times 10^{13}$ cm$^{-2}$) introduced by Boiss\'{e} et al. (1998).
Therefore, most previous surveys (e.g. Kulkarni \& Fall 2002; Prochaska et
al. 2003) suggest the mean metallicity measured by DLAs is
$\langle Z \rangle \sim -1.5$ at a redshift $z \sim 2$, which is significantly
lower than the metallicity $Z > -1.0$ measured in the Lyman break galaxies at
the similar redshift (Giavalisco 2002; Fynbo et al. 2008) or other starlight
emission-selected galaxies. This
superficial ``missing metals problem'' might be solved by taking into
account the significant population of dust obscured DLAs bearing bulks of
heavy elements. Note that the metallicity of J1156$+$6155, $Z \sim -0.6$, is
consistent with the measurement in emission-selected galaxies at
$z \sim 2$. The dusty DLA population might be an important metal reservoir
in the high redshift universe.

DLA J1157$+$6155 has a MW-like extinction curve with 2175~{\AA} dust
extinction bump and MW-like dust depletion patterns and MW-like metallicity.
The three features suggests that its counterpart galaxy may be an analog of the
MW in the high redshift universe. Jiang et al. (2010b, 2011) detected three dozens of
2175-{\AA} quasar absorbers with redshift of $z \sim 1.5$ in SDSS DR3.
Most of these dusty quasar absorbers have strong metal absorption lines and thus
may be metal-rich DLAs. In this paper, we have developed a more sensitive method
to detected 2175~{\AA} bump on quasar spectra. We plan to apply the new method to
the quasars in SDSS DR9\footnote{http://www.sdss3.org/dr9/} and expect to
find more than 100 dusty absorbers with significant 2175~{\AA} extinction bump.

\acknowledgments
We would like to thank the anonymous referee for
the enlightening suggestions that greatly improved the paper. We
thank Yueheng Xu for helpful comments on the manuscripts. This work
is supported by the Chinese NSF grant 10973012, 11033007 and 10973034, the SOC
program CHINARE2012-02-03 and the National Basic Research Program of
(973 Program) 2009CB824800. This work is also partially supported by
NSF with grant NSF AST-0451407, AST-0451408 \& AST-0705139 and the
University of Florida. This research has also been partially
supported by the CAS/SAFEA International Partnership Program for
Creative Research Teams. J. X. P. is supported by NSF grant
(AST-0709235).

This work has made use of data obtained by the SDSS and MMT. The
MMT telescope is operated by the MMT observatory, a joint venture
of the smithsonian institution and the University of Arizona.
Funding for the SDSS and SDSS-II has been provided by the Alfred P.
Sloan Foundation, the Participating Institutions, the National
Science Foundation, the US Department of Energy, the National
Aeronautics and Space Administration, the Japanese Monbukagakusho,
the Max Planck Society, and the Higher Education Funding Council for
England. The SDSS Web Site is http://www.sdss.org/.

\clearpage

\begin{figure}[tbp]
\label{fig-1}\epsscale{1} \plotone{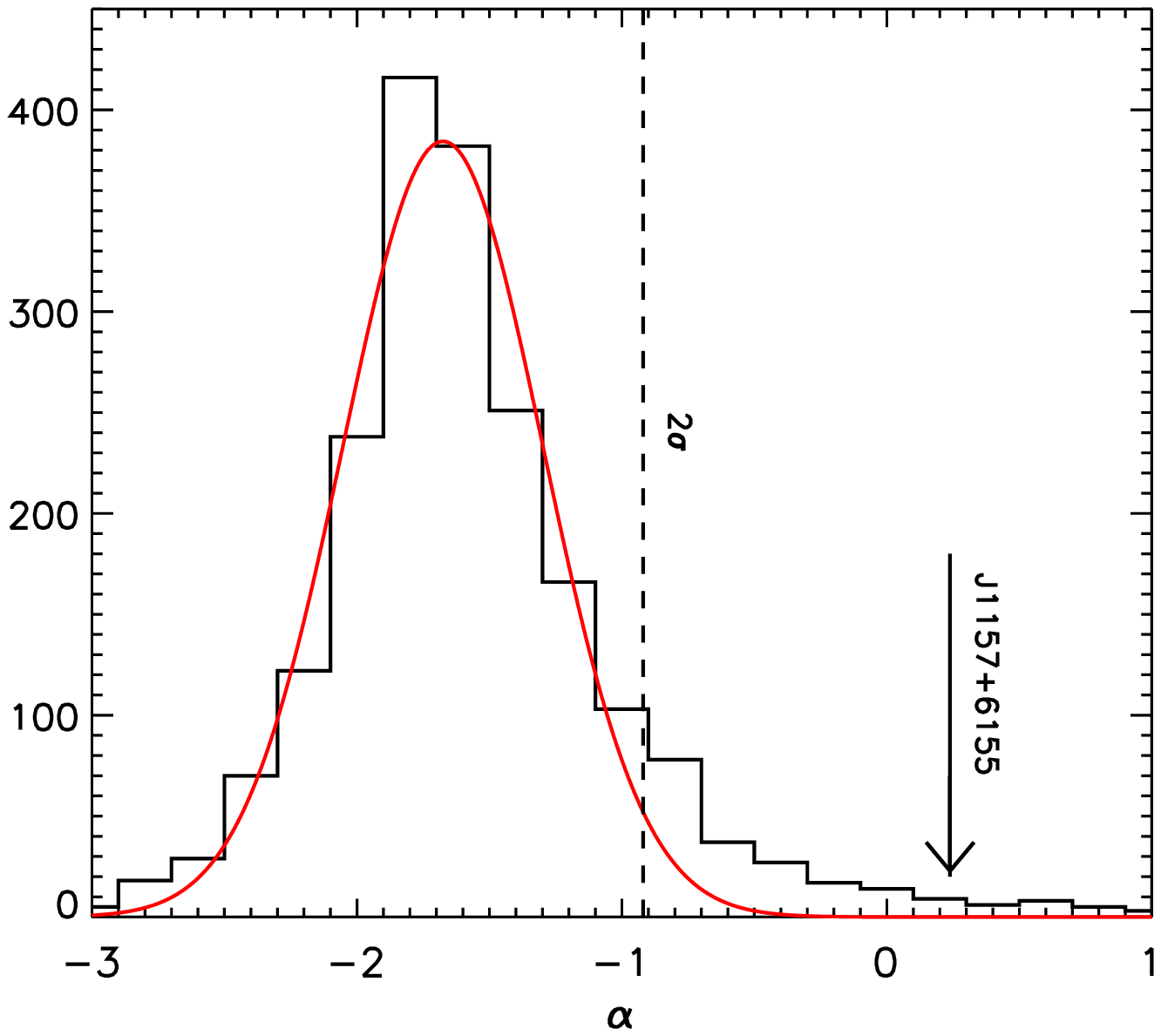} \caption{
Distribution of the spectral indices $\alpha$ (defined as
$S_{\lambda}\propto \lambda^{\alpha}$) of quasars from the SDSS DR7
in the redshift interval $2.4 \le z \le 2.6$.
Overlayed is a Gaussian curve fitted to the data.
The vertical dashed line indicates the $2\sigma$
deviation from the center of the distribution.
The value for J1157$+$6155 is indicated
by the arrow.}
\end{figure}
\clearpage

\begin{figure}[tbp]
\label{fig-2}\epsscale{0.8} \plotone{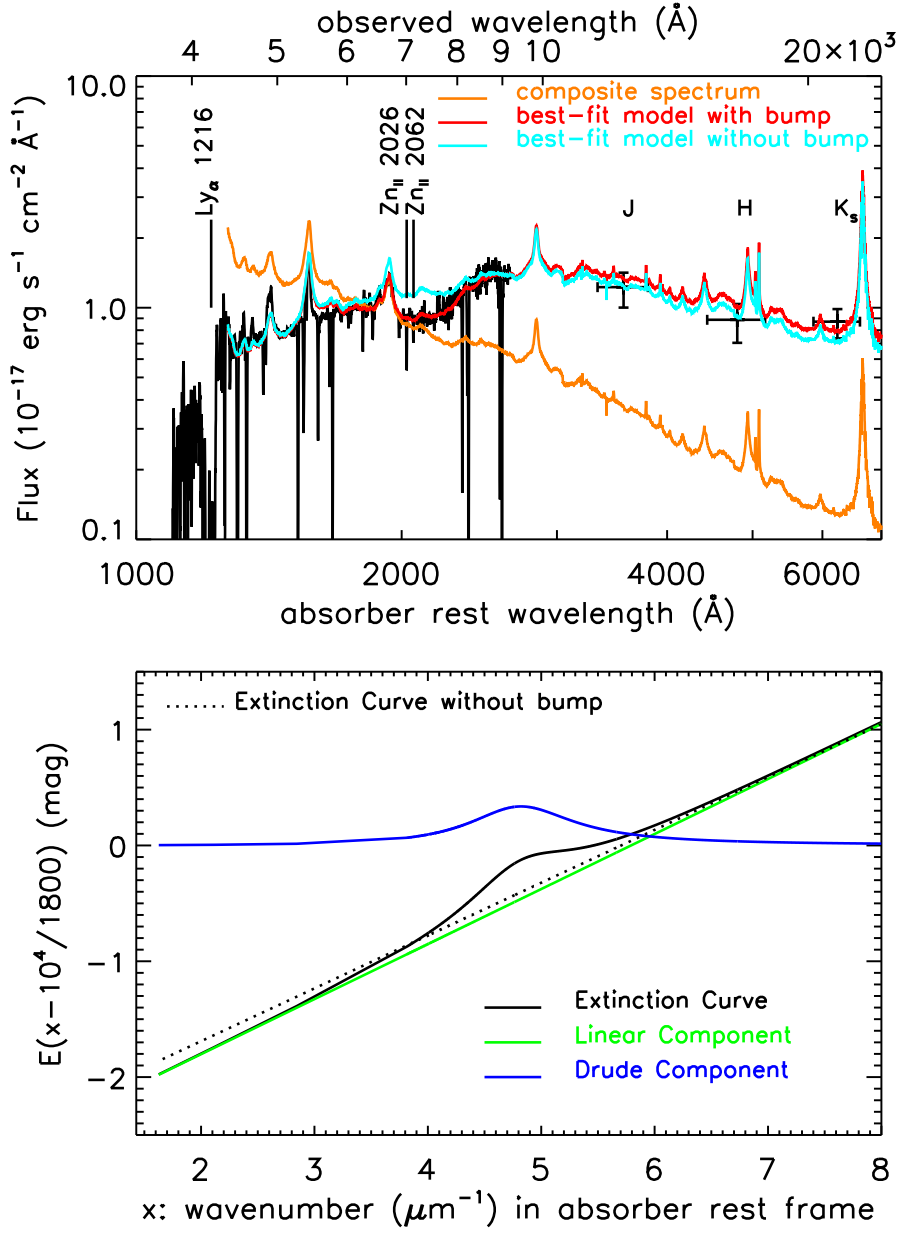} \caption{Upper: the
observed data of J1157$+$6155. The observed spectrum is a
combination of the MMT and SDSS spectra
(black, smoothed by a 7-pixel boxcar).
The composite quasar spectrum (orange) and the
best-fit reddened model (red) are overplotted with the observed one.
The 2MASS measurements in the
$J$, $H$ and $K_s$ bands are consistent with the best-fit model.
The cyan curve shows the model by reddening the
composite quasar spectrum with a SMC-like extinction curve
without a bump at around 2100 \AA\ in the DLA rest frame.
Lower: The best-fit reddening curve (black
solid) and its composites (green: a linear component; blue:
a drude component).
The fitted reddening curve without the bump was also
plotted for comparison (dotted).}
\end{figure}
\clearpage

\begin{figure}[tbp]
\label{fig-3}\epsscale{1} \plotone{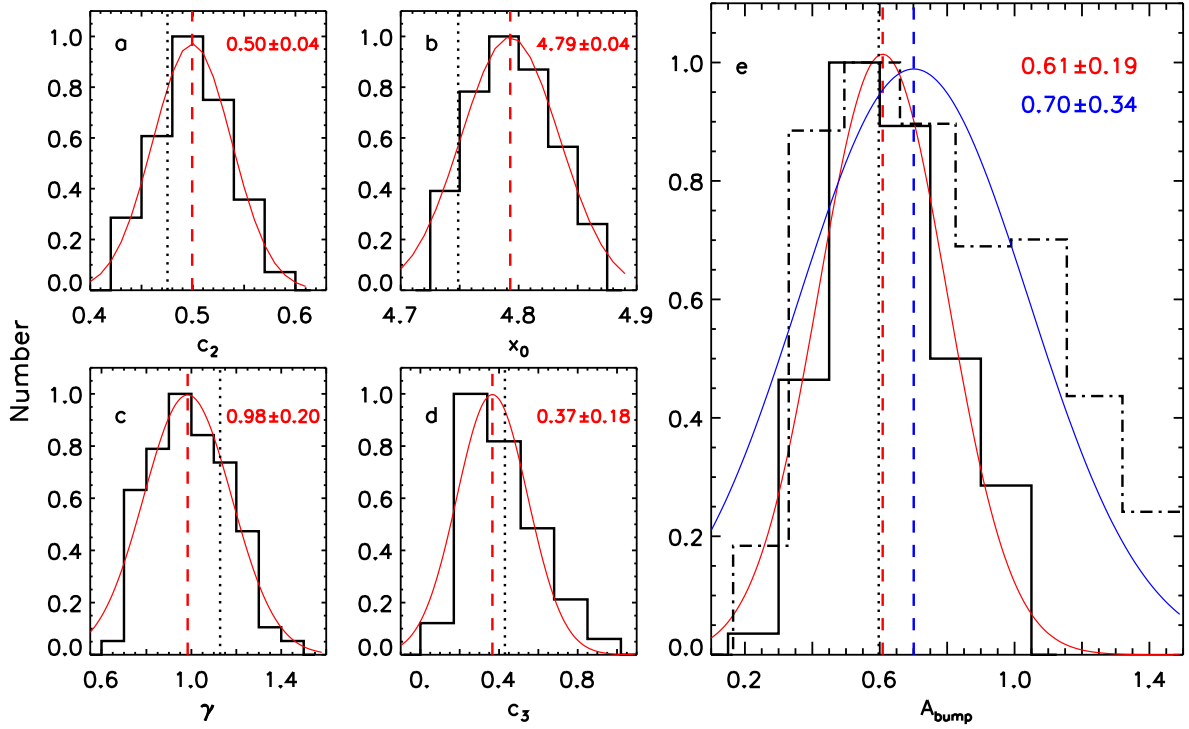} \caption{Panels $a,b,c,d$:
Distributions of the slope $c_2$, bump centroid $x_0$, bump width
$\gamma$ and parameter $c_3$ of the extinction curve estimated from
\feii\ matched template spectra (solid histogram) and a Gaussian fit to them (red).
The red dashed vertical lines show the best-fit results. The dotted
vertical lines show the best-fit results of the composite quasar spectra method.
Panel $e$: Comparison of the bump strengths derived with the \feii\ matched template
spectra (in red) and with the full template library (in blue). The blue Gaussian is nearly
twice broader than the red Gaussian, indicating that the refined modeling
procedure is necessary for detecting relatively weak 2175~{\AA} absorption
bumps on quasar spectra. See the text for details.}
\end{figure}
\clearpage

\begin{figure}[tbp]
\label{fig-4}\epsscale{0.8} \plotone{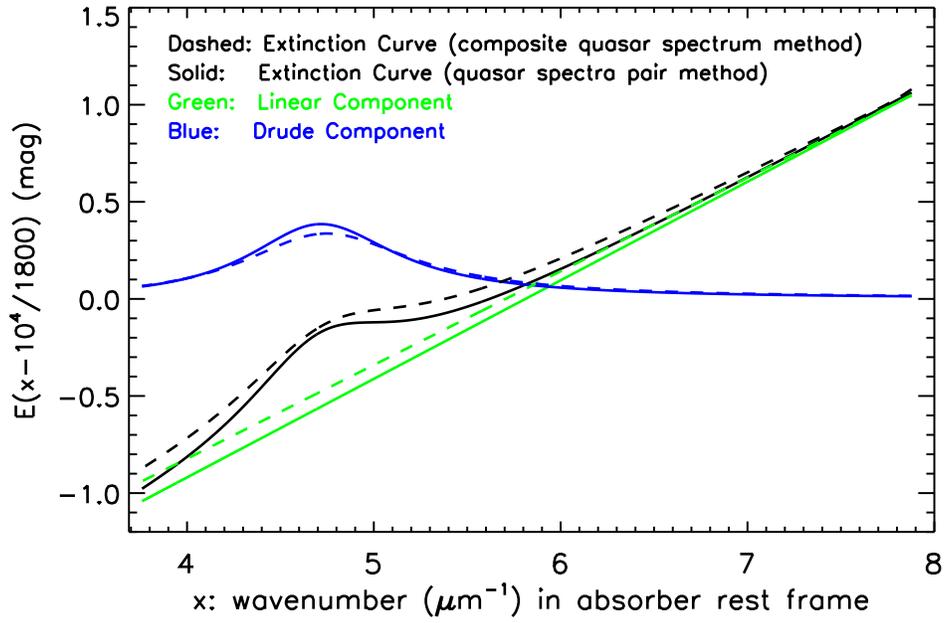} \caption{
The comparison of extinction curve derived with quasar spectra pair method (solid)
and that of the composite quasar spectrum method (dashed).}
\end{figure}
\clearpage

\begin{figure}[tbp]
\label{fig-5}\epsscale{1} \plotone{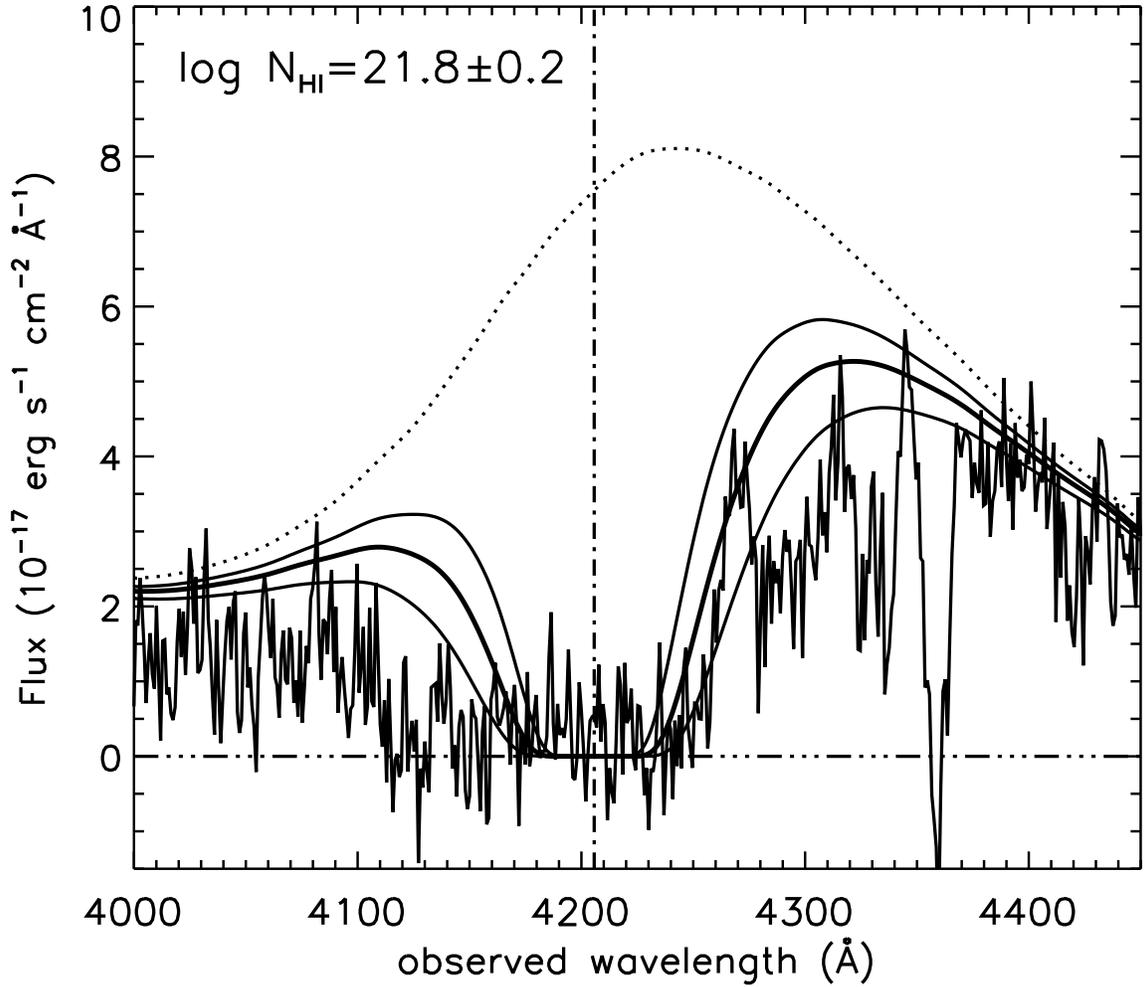} \caption{DLA line profile fitting on the
MMT spectrum (black). The vertical dash-dotted
line indicates the redshift estimated from narrow metal absorption lines.
The dotted line is the predicted model of the quasar spectrum.
The best-fit model is shown by the thick solid line, and the estimated 1 $\sigma$
uncertainty is given by the thin solid lines.
}
\end{figure}
\clearpage

\begin{figure}[tbp]
\label{fig-6}\epsscale{1} \plotone{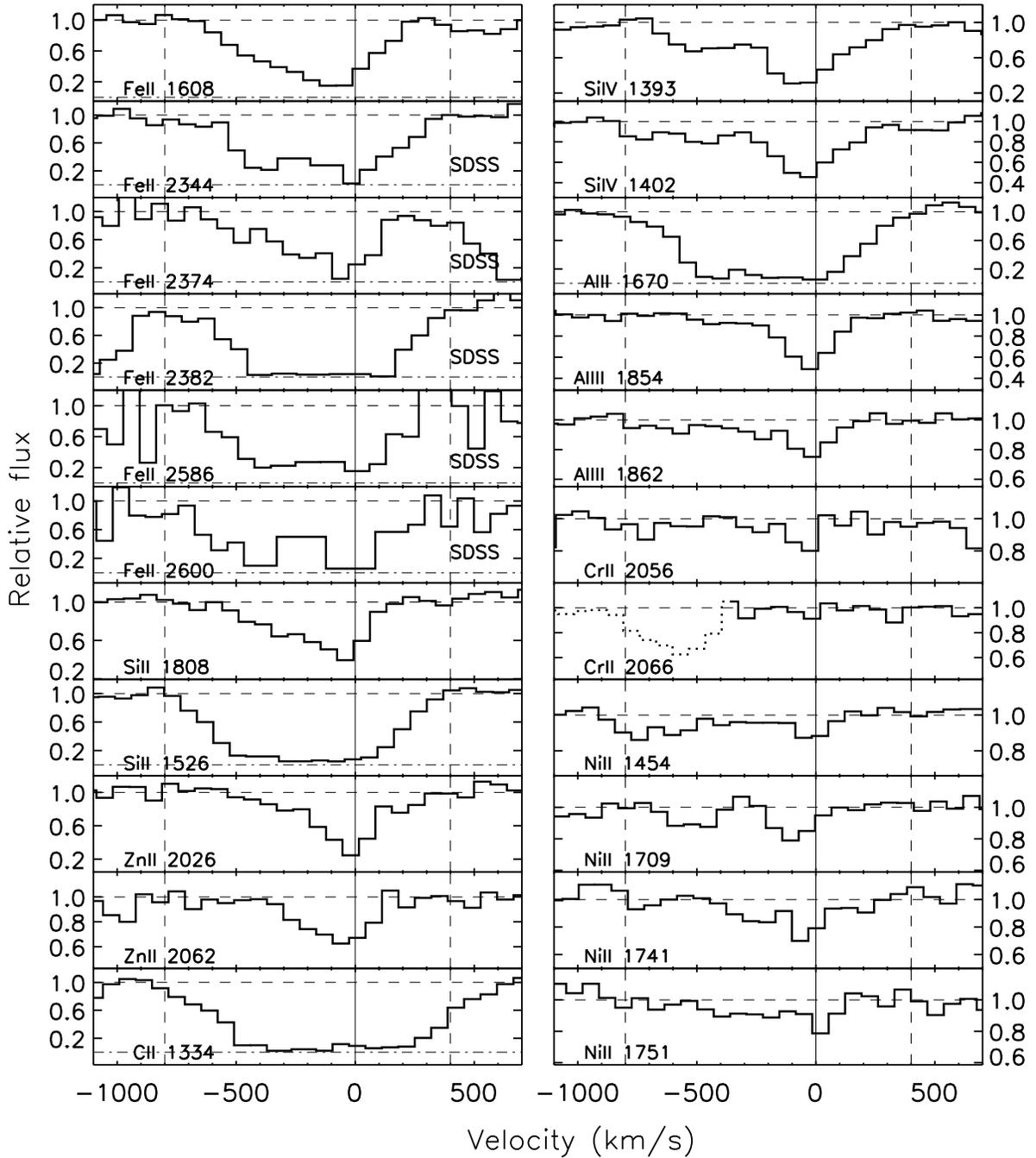} \caption{Normalized
absorption spectrum of J1157+6155.
The metal absorption lines are plotted in the velocity space.
The zero velocity point (the solid line) is corresponding to $z_{abs}=2.4596$.
The two vertical dashed lines show the
velocity interval used to measure the EW and $N_{AODM}$.
The Fe$\lambda\lambda$ 2344, 2374, 2382, 2586, 2600 absorption lines
are taken from SDSS spectrum
because they are not covered by MMT spectrum, while the rest of the lines
are measured from the MMT spectra.}
\end{figure}
\clearpage

\begin{figure}[tbp]
\label{fig-7}\epsscale{1} \plotone{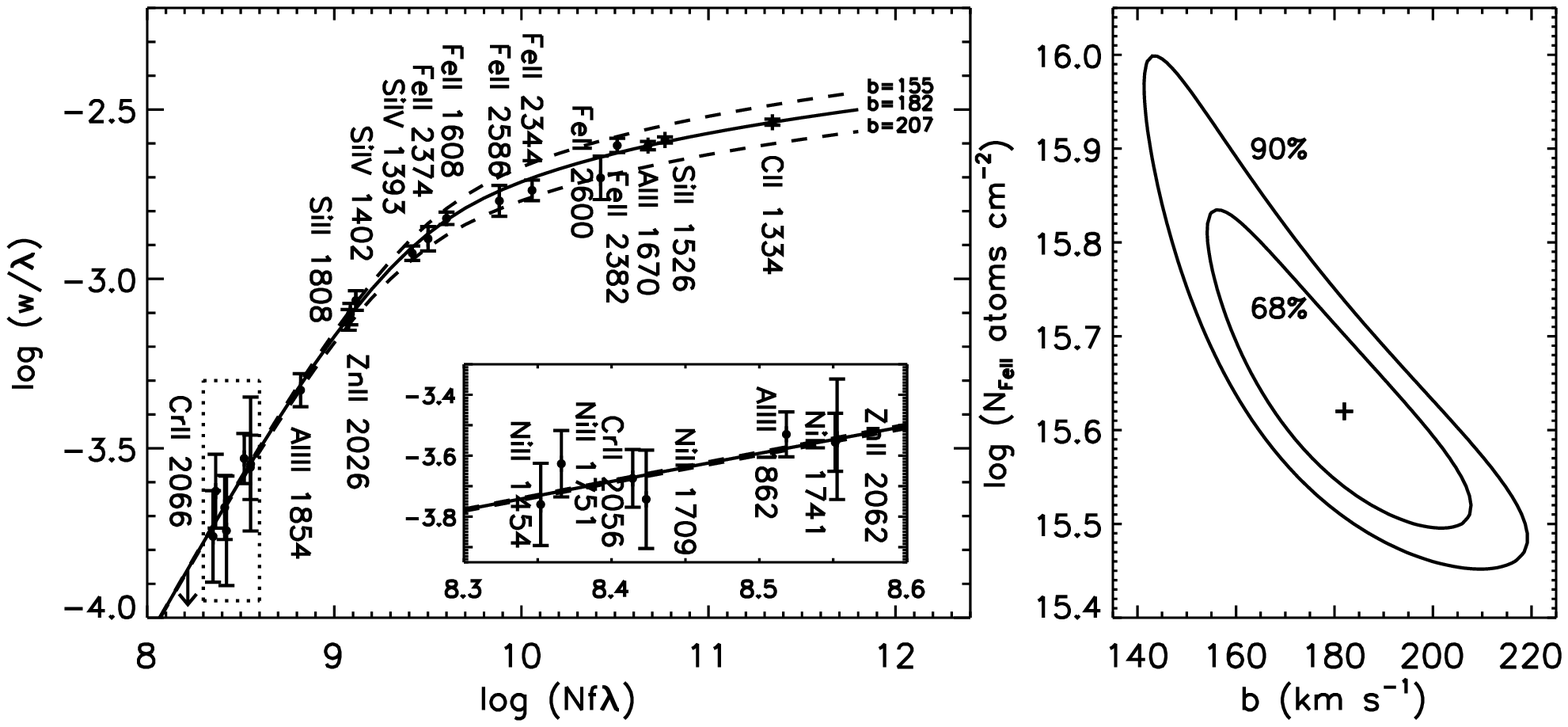} \caption{Left panel:
Best-fit COG using six Fe{\scriptsize II} absorption lines. The thick
solid line represents the best-fit curve and
the two dashed lines
show the boundaries of the 68\% confidence interval, as shown in the right panel.
The dotted square is enlarged in the inserted.
Right panel: $\chi^2$ contours in the $\log N_{Fe{\scriptsize II}}$-b
plane. The minimum $\chi^2$ is marked as a cross.}
\end{figure}
\clearpage

\begin{figure}[tbp]
\label{fig-8}\epsscale{1} \plotone{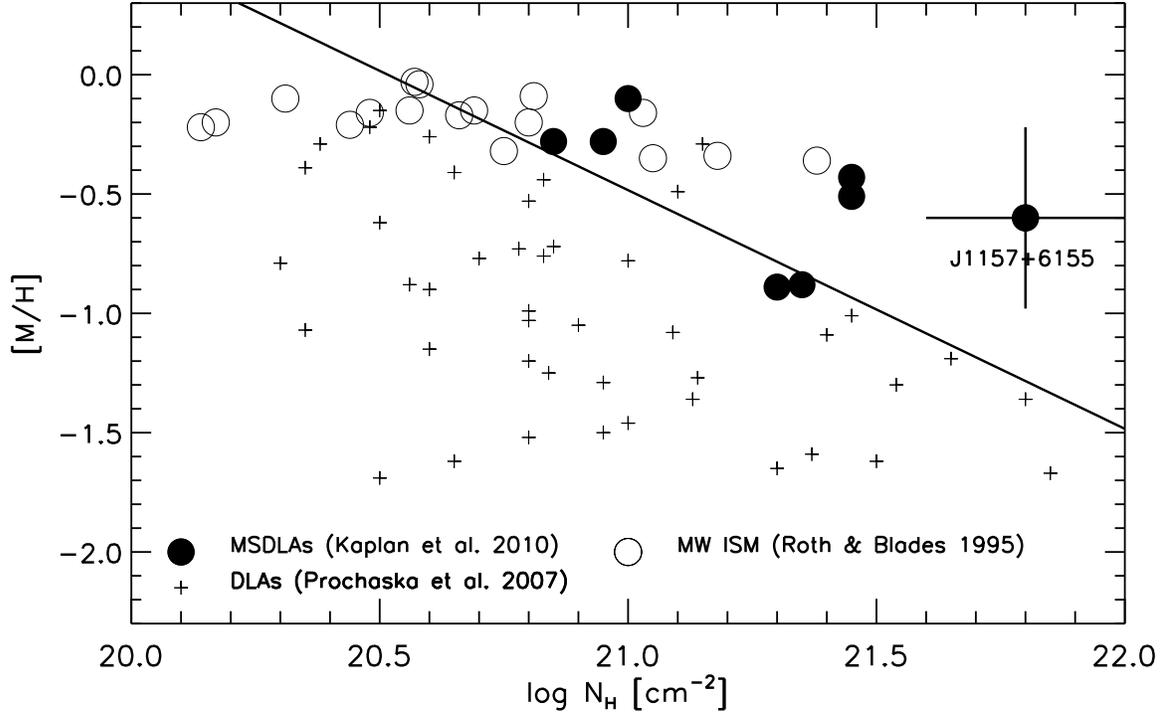} \caption{
Metallicity in different absorbers. The tiny plus symbols
are the normal DLAs in Prochaska et al. (2007). The [M/H]
of these objects are measured from $Zn^+$, $Si^+$ or $S^+$.
Nearly all of them are sitting under the dust obscuration threshold
(the solid line) defined by Boiss\'{e} et al. (1998). The open
circles are the clouds in the MW, where the metallicity is
measured by [Zn/H] (Roth \& Blades 1995). The metal-strong
DLAs (Kaplan et al. 2010), including J1157$+$6155 in this work,
are labeled as filled circles. Most of them are located above
the obscuration threshold and have similar metallicity
([Zn/H]) with MW clouds. Especially, J1157$+$6155 have the highest column
density of neutral hydrogen among the known metal-strong DLAs.
}
\end{figure}
\clearpage

\begin{figure}[tbp]
\label{fig-9}\epsscale{1} \plotone{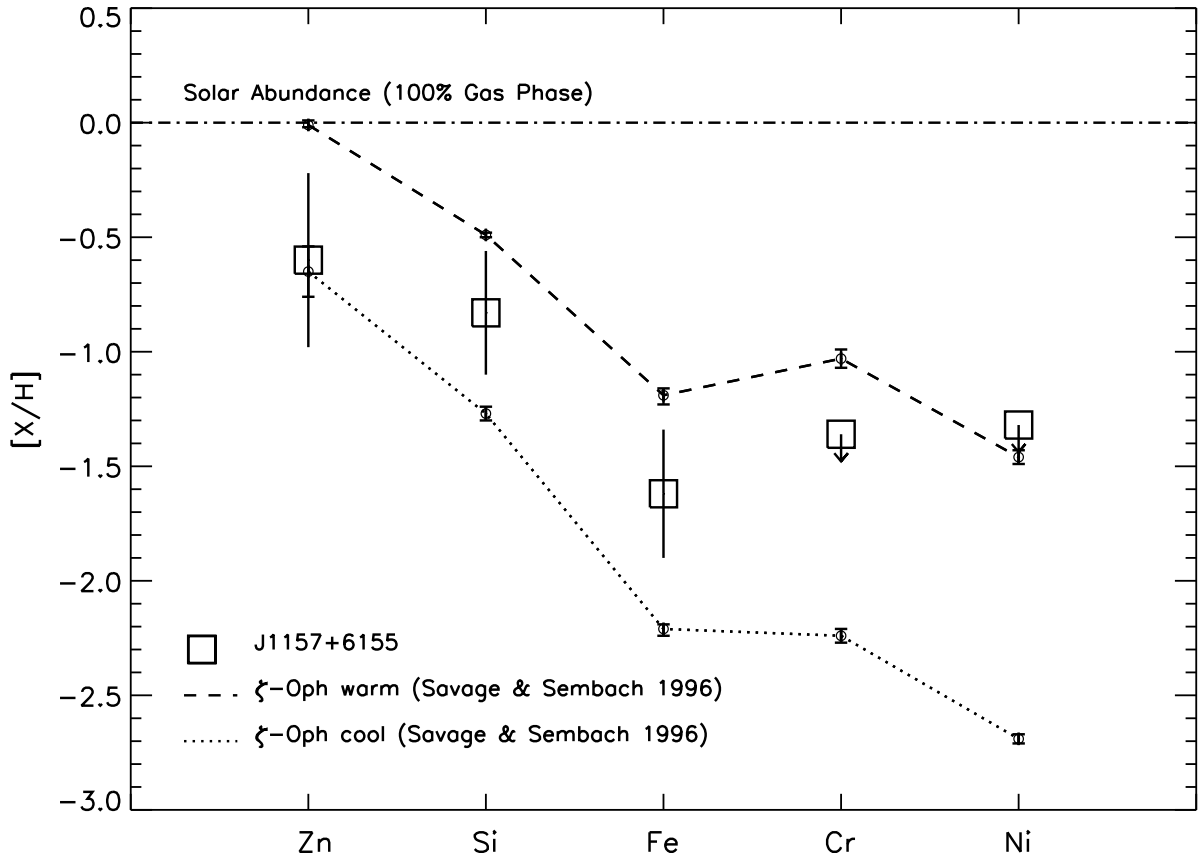} \caption{
The dust depletion pattern of J1157$+$6155 is similar
with the pattern measured in the MW, lying between that
of the warm and cool clouds in the sight line toward $\zeta$ Oph.}
\end{figure}
\clearpage

\begin{figure}[tbp]
\label{fig-10}\epsscale{1} \plotone{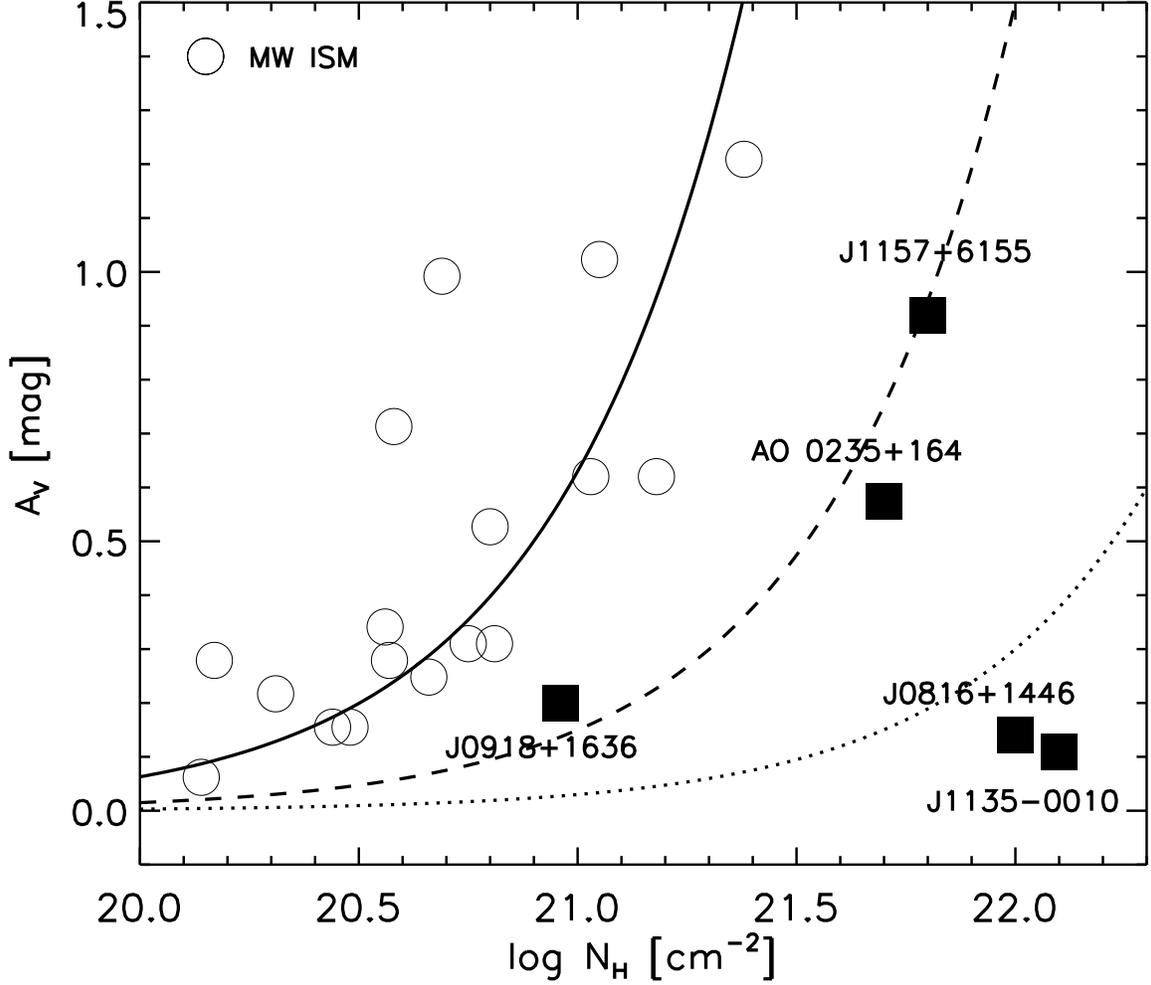}
\caption{Diversity of the dust-to-gas ratio of
DLAs. The dust-to-gas ratio of MW clouds
are shown as open circles, the average value of
which is $A_V = 6.3\times 10^{-22} N_H$ (the solid line; Diplas \& Savage 1994).
The average dust-to-gas ratio of DLAs, $A_V = 3\times 10^{-23} N_H$
(the dotted line; Vladilo et al. 2008), is much smaller than that of MW clouds.
Moreover, the dispersion of dust-to-gas ratios of DLAs
is quite large. For example, the two super-DLAs
(J0816$+$1446 and J1135$-$0010) have high column densities of
gas but very little dust content. While the dust-to-gas ratio (the dashed line) of the
three dust-rich
DLAs (J0918$+$1636 and AO 0235$+$164 as well as J1157$+$6155 in this work)
is larger than the average ratio by a factor of 5.
The large extinction of J1157$+$6155, $A_V \sim 1.0$, has been seen in the
most dusty MW clouds.
}
\end{figure}
\clearpage

\begin{deluxetable}{lccllc}
\centering
\tabletypesize{\scriptsize}
\tablenum{1}
\tablecaption{Ionic Column Densities for DLA J1157$+$6135}
\tablehead{\colhead{Ion} & \colhead{Transition} & \colhead{$\log f$} &
\colhead{EW} & \colhead{$\log N_{AODM}$} & \colhead{$\log N_{COG}$} \\
\colhead{} & \colhead{(\AA)} & \colhead{} & \colhead{(\AA)} &
\colhead{(cm$^{-2}$)} & \colhead{(cm$^{-2}$)}}
\startdata
Fe\,II  & 1608.4510 & -1.2366 & $2.43\pm0.14$   & $15.49\pm0.10$   & $15.63\pm0.19$ \\
        & 2344.2141 & -0.9431 & $4.28\pm0.33$   & $15.16\pm0.11$   & ... \\
        & 2374.4612 & -1.5045 & $3.12\pm0.29$   & $15.55\pm0.12$   & ... \\
        & 2382.7649 & -0.4949 & $5.90\pm0.31$   & $14.93\pm0.13$   & ... \\
        & 2586.6499 & -1.1605 & $4.40\pm0.50$   & $15.27\pm0.13$   & ... \\
        & 2600.1729 & -0.6216 & $5.17\pm0.83$   & $14.74\pm0.15$   & ... \\
Si\,II  & 1808.0129 & -2.6603 & $1.37\pm0.14$   & $16.45\pm0.10$   & $16.48\pm0.18$ \\
        & 1526.7065 & -0.8962 & $3.93\pm0.13$   & $15.58\pm0.11$   & ... \\
Zn\,II  & 2026.1360 & -0.3107 & $1.59\pm0.16^a$ & $14.10\pm0.11^a$ & $14.09\pm0.11$ $^a$\\
        & 2062.6641 & -0.5918 & $0.91\pm0.15^b$ & $14.04\pm0.11^b$ & $13.83\pm0.33$ $^b$\\
C\,II   & 1334.5323 & -0.8935 & $3.87\pm0.13$   & $15.69\pm0.11$   & $17.11\pm0.70$ \\
Si\,IV  & 1393.7550 & -0.2774 & $1.66\pm0.13$   & $14.40\pm0.10$   & $14.55\pm0.12$ \\
        & 1402.7700 & -0.5817 & $1.21\pm0.13$   & $14.52\pm0.10$   & ... \\
Al\,II  & 1670.7874 &  0.2742 & $4.14\pm0.15$   & $14.32\pm0.11$   & $15.18\pm0.49$ \\
Al\,III & 1854.7164 & -0.2684 & $0.87\pm0.14$   & $13.81\pm0.11$   & $13.82\pm0.12$ \\
        & 1862.7896 & -0.5719 & $0.55\pm0.14$   & $13.86\pm0.12$   & ... \\
Cr\,II  & 2056.2539 & -0.9788 & $0.44\pm0.15$   & $14.07\pm0.15$   & $<14.08$ $^c$\\
        & 2066.1609 & -1.2882 & $0.16\pm0.14$   & $13.82\pm0.35$   & $<14.19$ $^c$\\
Ni\,II  & 1454.8420 & -1.4908 & $0.25\pm0.14$   & $14.70\pm0.16$   & $<14.68$ $^c$\\
        & 1709.6042 & -1.4895 & $0.31\pm0.17$   & $14.61\pm0.20$   & ... \\
        & 1741.5531 & -1.3696 & $0.48\pm0.16$   & $14.70\pm0.13$   & ... \\
        & 1751.9156 & -1.5575 & $0.41\pm0.15$   & $14.73\pm0.16$   & ... \\

\enddata
\tablecomments{Equivalent width is measured in the absorber's
rest frame at $z=2.4596$. The vacuum wavelengths
and oscillator strength $f$ are adopted from the Atomic Data
compiled by J. X. Prochaska
(http://kingpin.ucsd.edu/$\sim$hiresdla/atomic.dat).
All statistical uncertainties represent the
68\% confidence interval.
The errors of $N_{AODM}$ include a systematic error of 0.10 dex
due to the uncertainty of normalization.
$N_{COG}$ is derived by using the best-fit COG and
assuming the same column densities of different
transition of same ions;
the errors include the errors caused by the uncertainties in EWs and Doppler parameter b,
as well as a systematic error of 0.10 dex due to the uncertainty of normalization.}
\tablenotetext{a}{The contamination of Mg\,I 2026 and Cr\,II 2026 are not removed.}
\tablenotetext{b}{The contamination of Cr\,II 2026 is
not removed from EW and $N_{AODM}$. When calculating $N_{COG}$, the contamination of
Cr\,II 2062 is removed by scaling the EW of Cr\,II 2056 according to their oscillator
strengths.}
\tablenotetext{c}{Derived from the $3\sigma$ upper limit of the EW.}
\end{deluxetable}

\end{document}